# Observation of Electromagnetic Fluctuation Induced Particle Transport in ETG Dominated Large Laboratory Plasma

Prabhakar Srivastav[1,2], Rameswar Singh[3], L. M. Awasthi[1,2], A. K. Sanyasi[1], P. K. Srivastava[1], Ritesh Sugandhi[1,2] and R. Singh[4]

[1]*Institute for Plasma Research, Gandhinagar 382428, India.*

[2]*Homi Bhabha National Institute, Mumbai 400085, India.*

[3]*University of California San Diego, United States.*

[4]*Advance Technology Center, NFRI, Rep. Korea.*

## Abstract

The Large Volume Plasma device(LVPD), a cylindrically shaped, linear plasma device of dimension ($\phi = 2m, L = 3m$) have successfully demonstrated the excitation of Electron Temperature Gradient(ETG) turbulence[ Mattoo et al.,[1]]. The observed ETG turbulence shows significant power between $\omega \sim (25-90)\ krad/s > \Omega_{ci}(= 8\ krad/s)$, corresponding to the wavenumber, $k_y \rho_e \sim (0.1-0.2) < 1$ where, $\Omega_{ci}$ and $\rho_e (\approx 5\ mm)$, ion cyclotron frequency and electron Larmor radius, respectively. The observed frequency and wave number matches well with theoretical estimates corresponding to Whistler-ETG mode. We investigated electromagnetic (EM) fluctuations induced plasma transport in high beta ($\beta = 2\mu_o n T_e / B^2 \sim 0.01-0.4$) ETG mode suitable plasma in LVPD. The radial electromagnetic (EM) electron (ion) flux ($\Gamma_{em}^{e,i}$) results primarily from the correlation between fluctuations of parallel electron current ($\delta J_{\parallel,e,i}$) and radial magnetic field ($\delta B_r$). The electromagnetic particle flux is observed to be much smaller than the electrostatic particle flux, $\Gamma_{em} \approx 10^{-5} \Gamma_{es}$. The EM flux is small but finite contrary to the conventional slab ETG model. The electromagnetic flux is non-ambipolar. A theoretical model is obtained for the EM particle flux in straight homogeneous magnetic field geometry. The theoretical estimates are seen to compare well with the experimental observations. Sluggish parallel ion response is identified as the key mechanism for generation of small but finite electromagnetic flux.

# I. Introduction

Plasma transport continues to remain a core problem for magnetically confined fusion plasmas[2,3] and hence motivates continued efforts in the domain of theoretical, computational and experimental investigations for a better understanding of the physical phenomena responsible for it. The observed plasma loss is found to be an order higher than the predicted classical and neo-classical values and is attributed to the turbulent transport caused by the fluctuations. These fluctuations can have electrostatic and or electromagnetic nature. Hence the net turbulent particle flux is made of the sum of electrostatic and electromagnetic pieces, $\Gamma = \Gamma_{es}^q + \Gamma_{em}^q$, where $\Gamma_{es}$ and $\Gamma_{em}$ represents the electrostatic and electromagnetic flux pieces. The radial electrostatic flux appears due to correlation between density, $\tilde{n}$ and perpendicular electric field fluctuations, $\tilde{E}_\perp$ i.e., $\Gamma_{es} = <\tilde{n}\tilde{E}_\perp>/B$. The radial electromagnetic flux forms from correlation between parallel current, $\delta J_{\parallel,q}$ of charge species, $q \approx e, i$ and radial magnetic field fluctuations, $\delta B_r$. Although significant advancement in developing an understanding is achieved as far as electrostatic flux contribution is concerned[4–8] the electromagnetic fluctuations driven radial flux is still not completely explored. Although, the electromagnetic particle flux has been reported from many devices, namely, the tokamaks[9,10], the stellerator[11] and reversed field pinch [12–14] etc. but despite this, electromagnetic electron scale fluctuations induced particle transport remain largely unaddressed. Despite extensive theoretical and numerical efforts on Electron Temperature Gradient(ETG) turbulence for tokamaks, the measurement of particle and energy transport due to ETG turbulence remain unexplored. Indirect inferences of its existence has been reported by various tokamaks viz., ASDEX[15], Tore Supra[16] etc. but direct measurement of it has been reported only from NSTX[17,18]. In the recent times, some of the low temperature laboratory plasma devices have reported successful demonstration of ETG[1,19,20]. The advantage these devices carry unlike their high temperature counterparts is their ease of operation and control of parameters which has helped them in reporting particle and energy transport in ETG background, in particular due to the electrostatic fluctuations. While the effect of the electromagnetic fluctuations on particle transport still remain unaddressed, the streaming of charged particles parallel to fluctuating magnetic field is considered as a powerful transport mechanism, particularly if field lines wander stochastically in space[21,22]. The significance of magnetic fluctuations in the edge of various toroidal devices suggests that they are very significant in contributing to transport in different device configuration regimes like reversed field pinches[23], high beta tokamaks[24], and tokamaks in L-H transition[25].

The magnetic fluctuations can be driven by different instabilities present in plasma due to inhomogeneity in plasma parameters and basic understanding of magnetic fluctuation induced particle transport processes is thus of great interest and potentially critical to plasma density control and understanding fast particle losses in future plasmas like ITER. These energetic particles are present in magnetic fusion devices due to external plasma heating and eventually due to fusion born α- particles. It is necessary that these super- thermal particles are well confined while they transfer their energy to the background plasma[26]. As reported, magnetic fluctuations induced plasma transport can be large even if the fluctuation amplitude is extremely small[27]. The excited magnetic field fluctuations in high-temperature machines have their origin lies with instabilities excited during non-inductive heating and current drive by energetic particles. Even, they get excited by global tearing instabilities that often underlie the saw tooth oscillation associated with magnetic reconnection and lead to plasma relaxation. The stochastic magnetic field is deliberately imposed in tokamak plasmas by using external coils to mitigate the edge-localized modes excited electromagnetic particle flux[28]. However, transport due to magnetic fluctuations has been mainly studied indirectly by measuring runaway electron flux to a limiter. Such experiments are useful for probing the magnetic fluctuations but do not provide a local measurement of particle transport resulting from the magnetic field. In the past, direct measurement of magnetic fluctuation induced particle flux indeed has been reported but has been limited to the edge region of fusion devices using probes[13,27]. However, not much progress was made for the core plasma where the temperature is high due to the inability of measurement tools.

In the recent time, Ding et al.[12], have demonstrated measurement of magnetic field fluctuation induced particle transport from the core plasma of Madison RFP device. The presence of stochastic magnetic field supports non ambipolar transport (charge transport) since electrons stream rapidly along field lines. In such a situation, they report that measured electron flux is responsible for the rapid particle transport and it modifies the equilibrium density[29]. They also reported that the measured particle transport exceeds the expected particle diffusion when ambipolarity is maintained because of the slowing down of an electron to ion diffusion but it agrees well with the expected values when ambipolarity is absent. In a theoretical sense, non-ambipolarity can exist only when it is balanced by an opposing non-ambipolar flux in order to maintain plasma quasi-neutrality. However, Stoneking et al.[27], Rempel et al.[13], and Shen et al.[28] have reported that particle loss induced by magnetic field fluctuations is ambipolar in the edge plasma.

In high beta plasma of LVPD ( $\beta \sim 0.2$ ), the magnetic field fluctuations couples with the ETG mode instability to form what is dubbed as a whistler-ETG mode. In low beta plasma, the slab ETG mode is primarily driven by parallel compression of electron motion along the magnetic field. The compression effect in electron parallel motion will generate temperature and density perturbations. The density perturbation is an out of phase to potential perturbation via ion Boltzmann shielding effect. This potential perturbation creates E x B drift, which brings cold electrons in a compressed region and thus lowers the pressure. The lower pressure attracts more electrons, further increasing the compression. This positive feedback loop leads to instability. However, in high beta plasma, Sushil et al.[30], have shown that when electromagnetic effects are included, the ion-electron interchange symmetry breaks down. This is because the magnetic perturbations alter the electron dynamics; as a result, the parallel dynamics of ions and electron are no longer symmetric. The coupling of Whistler and ETG mode becomes important when the beta of plasma is high (i.e.,$\beta \sim 0.01 - 0.4$) but W-ETG mode again becomes unstable like ETG only when the electron temperature gradient crosses a threshold value, $\eta_e > 2/3$. In magnetized plasma, the electromagnetic flux in ETG background is expected to be zero as per the conventional ETG mode theory, where it is assumed that the electron current fluctuation $\delta J_{\parallel,e}$ is the total current fluctuation, $\delta J_{\parallel}$. The radial electron particle flux can then be written as[31],

$$\Gamma_{em} = \frac{1}{eB} < \delta J_{\parallel e} \delta B_x > = \frac{1}{eB} < \delta J_{\parallel} \delta B_r > \quad = < \nabla_{\perp}^2 A_{\parallel} \frac{\partial}{\partial y} A_{\parallel} > = Real(\sum_{\vec{k}} i k_{\perp}^2 k_y \mid A_{\parallel} \mid^2) = 0$$
(1)

Where, $k_y$, $k_{\perp}$ are wave vectors in radial and perpendicular directions $and$ $A_{\parallel}$ is vector potential respectively.

Clearly, this vanishes because the summand is odd in poloidal wave number, $k_y$. Noticeably, we do observe a small (compared to electrostatic flux) but finite electromagnetic flux in the LVPD. Obviously this means that the assumption $\delta J_{\parallel e} = \delta J_{\parallel}$ is broken in reality and consequently leads to a finite flux. This paper provides a detailed measurement of electromagnetic particle flux across the radius of the LVPD for the first time and provides a general theory of the electromagnetic particle flux in ETG turbulence in a straight homogeneous magnetic geometry which explains well the experimental observations. A finite electromagnetic radial flux is shown to result from the sluggish and passive parallel ion velocity fluctuations resulting from the parallel force experienced by ions due to ETG fluctuations. The motion is sluggish because the ion velocity scales as $\frac{m_e}{m_i}v_{\parallel e}$ and passive because this small mass ratio for Argon plasma barely effect the linear features of the ETG mode. Lack of magnetization of ions fails to produce a fluctuating radial drift of ions, which leads to no turbulent ion flux making the flux non-ambipolar. The electron flux is made of pieces resulting from parallel ion current fluctuations and resulting from magnetic stresses. The piece of flux resulting from ion current fluctuations is exactly similar to electromagnetic ion flux had the ions been magnetized like the ITG turbulence. Due to this morphological similarity it is dubbed here as pseudo ion flux. Quasilinear estimate is obtained for the pseudo ion flux which agrees well with the experimental measurement.

The rest of paper is organized as follows; the experimental setup and diagnostics are described in section II. The experimental observations are discussed in section III. In section IV, we discuss the theoretical model and its comparison with experimental results. Finally, the summary and conclusion are provided in section V.

## II.  Experimental Setup

The experimental setup primarily consists of Large Volume Plasma Device (LVPD)[32], the cathode, large electron energy filter(EEF) and diagnostics. The directional probes are developed for the measurement of parallel current density due to electrons, $J_{\parallel,e}$ and ions, $J_{\parallel,i}$ respectively[33,34]. Miniature bi-filer B-dot probes are used for measuring the three components of magnetic field ($B_x, B_y, B_z$) along with other conventional tools, purposefully mounted in device from its different radial ports viz., Langmuir, emissive probes etc. for gathering information on basic plasma parameters [Figure 1]. The large volume plasma device is a cylindrical, double walled, water cooled, vacuum chamber of dimension ($\phi = 2m, L = 3m$)[32]. The source of primary ionizing electrons is a set of 36 filaments of dimension ($W, \phi = 0.5mm, L = 16cm, A_{emission} \sim 75cm^2$), which are deployed on the periphery ($90\,cm \times 130\,cm$) of a rectangular, water-cooled cusped plate. The $4kG$ Samarium Cobalt magnets are used to produce the cusp magnetic field. Axial confinement of plasma particles is provided by similar cusped end plates, placed axially opposite to the cathode plate [35]. The radial confinement is provided by a uniform 6.2 G axial magnetic field, produced by a set of 10 coils, garlanded over the surface of the vacuum chamber. The pulsed Argon plasma is produced of discharge duration ($\Delta t_{discharge} = 9.2 ms$) by applying a discharge voltage of $70V$ between the cathode and the vacuum chamber at a filling pressure of $P_{Ar} \sim 4 \times 10^{-4} mbar$. The device has a base pressure of $1.5 \times 10^{-6} mbar$.

In LVPD, we have produced plasma conditions suitable for carrying out an unambiguous investigation on ETG turbulence. The recipe for such plasma is characterized by uniform plasma density, sharp electron temperature gradient, and plasma devoid of non-thermal electrons. This ensures that excited plasma instabilities have origin only in electron temperature gradient. Usually, filamentary discharges contain a large population of non-thermal electrons and hence producing ETG suitable conditions is a tedious assignment. Also, establishing an independent control over density and temperature profiles is proved to be a difficult task. We dressed LVPD plasma to meet these requirements by inventing a large-sized electron energy filter[1,36] . The filter is a solenoid having rectangular cross section with 82% transparency and is placed across the diameter of the LVPD. It is of the variable cross-section with a maximum at its axis ($190 cm \times 4 cm$) and minimum near the walls ($4 cm \times 4 cm$). It divides LVPD plasma into three regions of source, EEF and target plasmas.

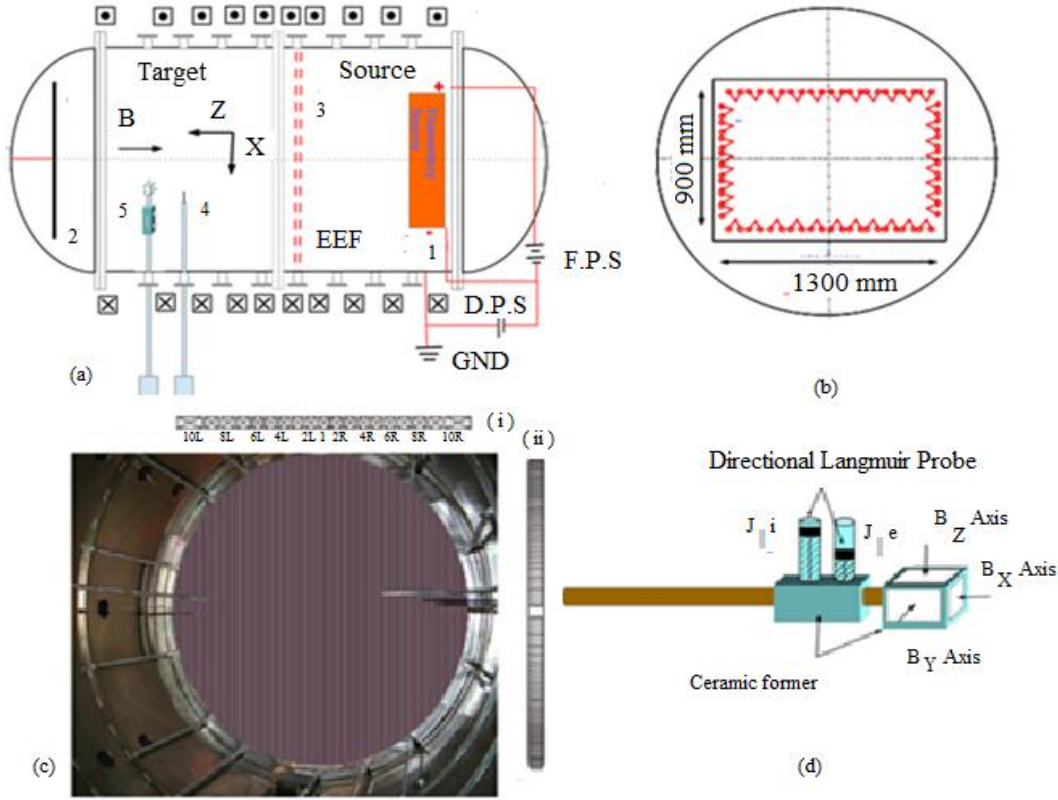

Figure 1: Schematic diagram of experimental setup is shown in (a), the layout shows internal components marked as (1) back plate, (2) end plate, (3) Electron Energy Filter(EEF), (4) Langmuir Probe, (5) a pair of B-dot and directional Langmuir probes for $J_{\parallel e,i}$ measurements, (b) cross-section view of LVPD showing the filament assembly arranged in a rectangular geometry (1300mm x 900 mm) in the source region, (c) the cross-sectional view of EEF in top horizontal bar, showing radial extent of each element of EEF and sidebar showing the respective spatial extent in vertical direction. Details of probe assembly (d) for EM flux measurement is marked as (5) in figure (a). This assembly contains a pair of directional Langmuir disc probes for parallel electron and ion current measurements and a 3-axis B-dot probe.

The source plasma embeds large population of energetic electrons, the EEF plasma is a low beta ($\beta \sim 10^{-3}$) sandwiched plasma between two high beta plasmas of the source ($\beta \sim 1.6$) and target ($\beta \sim 0.2$) regions[37]. The sandwiched EEF plasma region proposes a complex plasma scenario. We observed that energetic electrons are trapped in the mirror of the solenoid field of the EEF in the source plasma region of LVPD and no trace of these electrons is observed in the target region[36]. The transport across EEF magnetic field helps in producing a plasma profile suitable for the excitation of ETG instability, in axially far-off region, approximately $70\ cm$ away from the EEF surface.

A comparison of plasma parameters in the three regions is given in table 1. A detailed description of LVPD, EEF, information about characteristic features of different plasma regions and details of conventional diagnostics is already described in the referred work[32,36,37] . The present study focuses on the target region, where scale length of the gradient in density and electron temperature satisfies the ETG threshold conditions.

Table1: The typical plasma parameters obtained in the target, EEF and source plasmas of LVPD.

| Plasma Parameters | | | |
|---|---|---|---|
| | **Source** | **EEF** | **Target** |
| Plasma density, $n_e\ (cm^{-3})$ | $6.0 \times 10^{11}$ | $2.3 \times 10^{11}$ | $1 \times 10^{11}$ |
| Electron Temperature, $T_e(eV)$ | 5.0 | 2.5 | 2.2 |
| Plasma beta, $\beta$ | 1.6 | $10^{-3}$ | 0.2 |
| $f_{pe}(Hz)$ | $7 \times 10^9$ | $4.9 \times 10^9$ | $3.5 \times 10^9$ |
| $f_{pi}(Hz)$ | $3 \times 10^7$ | $1.8 \times 10^7$ | $1.3 \times 10^7$ |
| $f_{ce}(Hz)$ | $1.0 \times 10^7$ | $2.8 \times 10^8$ | $1.0 \times 10^7$ |
| $f_{ci}(Hz)$ | 236 | $6 \times 10^3$ | 236 |
| Electron-ion, $\nu_{ei}(s^{-1})$ | $\sim 1 \times 10^5$ | $\sim 4 \times 10^4$ | $\sim 1 \times 10^4$ |
| Debye length, $\lambda_D(cm)$ | $2.1 \times 10^{-3}$ | $2.3 \times 10^{-3}$ | $2.7 \times 10^{-3}$ |
| Electron gyroradius, $\rho_e\ (cm)$ | 0.8 | 0.02 | 0.5 |
| Ion gyro radius, $\rho_i(cm)$ | 73 | 2.2 | 46 |

Presently, for the investigations on EM flux estimation, we focused our attention on the measurement of parallel electron and ion currents along with different magnetic field components. For this purpose, we installed a specially designed probe array, accommodating a pair of directional probes for the measurement of fluctuations in parallel currents and a 3-axis B-dot probe for magnetic field fluctuation measurement. The directional probe assembly contains a disc probe of diameter, $\phi = 5\,mm$ and is encapsulated in a ceramic tube so that shielding restricts the solid angle through which probe collects particle and thus allows only the directed charge species. The pair of directional probe assembly, meant for parallel electron and ion current measurements are radially separated by $\Delta x = 8 mm$. Out of these two probes, the probe which is used for collecting parallel electron current is accommodated within the ceramic tube at a distance of 5mm [$\sim \rho_e$ (maximum at $x=0$)] thus allowing primarily collection of parallel electron current at all radial locations. The parallel electron current density $J_{\parallel e}$ is measured by keeping probe biased at plasma potential. Near plasma potential, probe collects electrons moving with electron thermal velocity, $V_{the} = \left(\frac{T_e}{m_e}\right)^{0.5}$. In the collision less scenario of LVPD, where electrons are magnetized, the parallel electron flux can be approximated as electron thermal flux, i.e. $J_{\parallel e} \approx J_{e,sat}$. Due to shielding of directional probe, only those electrons with Larmor radius, $\rho_e < \phi_{disc\,probe}$ and which are tied to field lines will contribute for parallel electron current density fluctuations. The electrons coming from random directions will be screened out. In the order to measure parallel electron current, we first established plasma potential measurement by making use of a center tapped emissive probe (CTEP)[38].

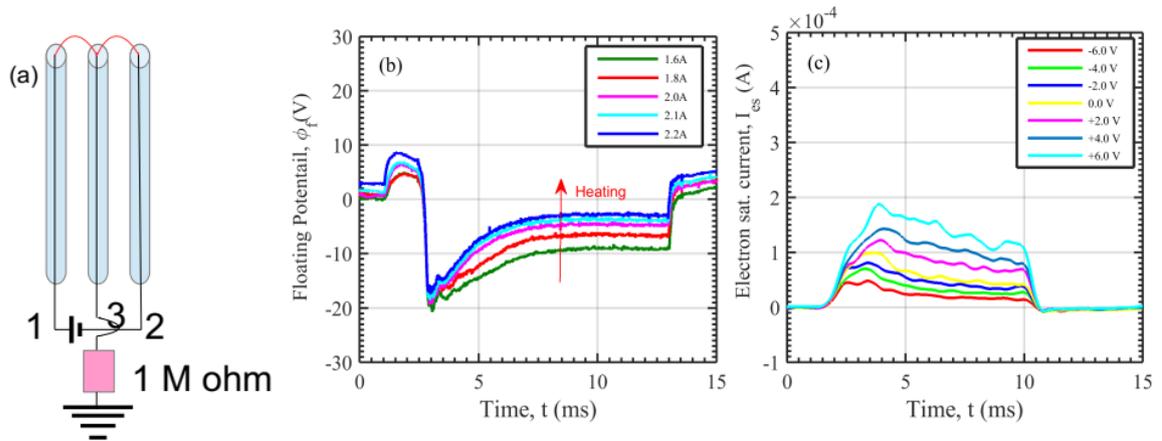

Figure 2: The schematic of the CTEP is shown in figure 2(a). In CTEP, floating potential is measured at nodal point 3, by heating it to different temperatures, operating CTEP in floating point technique. The floating potential is measured at the center of the device. The saturated CTEP potential at $I_{CTEP}$ ~2.1 give the measure of plasma potential, $\phi_P$ in (b). The parallel electron current measurements in (c) are made by keeping directional disc probe at potential $\geq$ $\phi_P$.

The center tapped emissive probe consists of a pair of emissive probes of equal electrical length and center of total electrical length serves as a center tap point. The three nodes of the probe are crimped to the gold plated beryllium copper pins. These pins are accommodated in a 3 bore ceramic tube and are isolated from plasma by the filled ceramic paste. The three bores of the ceramic tube holds all three nodes of the emissive probe. The two nodes 1 & 2 are elevated to potential –V and +V with respect to the common of DC power supply whereas the center tap point remain tied at zero potential (-V+V)/2=0). Measurements made at the junction remains unaffected due to the lifting of potential at two ends. A schematic diagram of CTEP is shown in 2(a).

The plasma potential is measured using CTEP by operating it using floating point technique. The CTEP heating current is varied till floating potential saturates. Figure 2(b) shows the saturation of floating potential at -4V for a heating current of ~ 2.1 A. This voltage is taken as bias potential for the measurement of parallel electron current. The bias potential to the disc probe, in the vicinity of plasma potential is varied till electron current approaches saturation. The typical variation of $I_{\|e}$ near to plasma potential is shown in figure 2(c). The estimation of parallel electron current is made at all radial locations by keeping the probe bias fixed at this voltage.

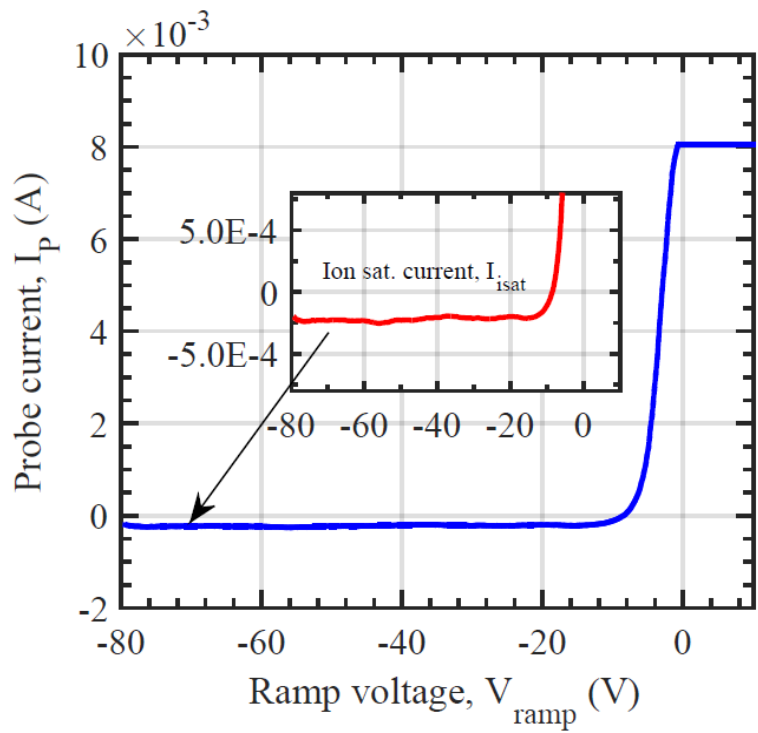

Figure 3: I/V curve obtained at x=0 for the directional disc probe used for measuring parallel ion current is shown. The almost saturated value of ion current indicates the negligible effect of sheath expansion in ion current measurements. The probe is biased at -80V to ensure contribution primarily for parallel ion current.

The second disc probe ($\phi_{disc} = 5mm$) of the array is used for measuring parallel ion current. This probe is kept at $2mm$ ($\gg 10\,\lambda_d$) inside the ceramic tube to cater the effect of ion sheath expansion.). The I/V curve obtained for this is shown in figure (3), suggesting that at high negative bias potential, the probe collects only ion current and its saturation indicates negligible effect of sheath expansion in ion current measurements. The current density is given as, $J_{i,sat} \approx 0.6\,n_i e c_s$, where $n_i$ is ion density, $c_s = \sqrt{\frac{T_e}{M_i}}$; Bohm velocity, and $e$ the electronic charge. The parallel ion current density is estimated following $J_{\parallel i} = en_i V_\parallel = (\frac{M}{0.6}) * J_{i,sat}$, where $\frac{V_\parallel}{c_s} \approx M$ [33,39,40].

The magnetic field fluctuations are measured using a 3 axis B-dot probe. This probe is configured in bi-filer geometry and is capable of measuring all three components of the magnetic field over the frequency band of (1-100) kHz. The design and construction are based on the Everson model of B-dot probe[41]. Its design eliminates electrostatic pickup, reduce physical size and increase the signal to noise ratio while maintaining a high bandwidth. The bifilar configuration ensures significant rejection of capacitive pickup which is excited due to the capacitive charging of probe during the plasma discharge. The probe is tested for capacitive rejection, inductive pickup and frequency response. An electrical equivalent of B-dot probe is shown in figure 4(a). The frequency response of B-dot probe in a uniform magnetic field of Helmholtz probe is shown in 4(b).

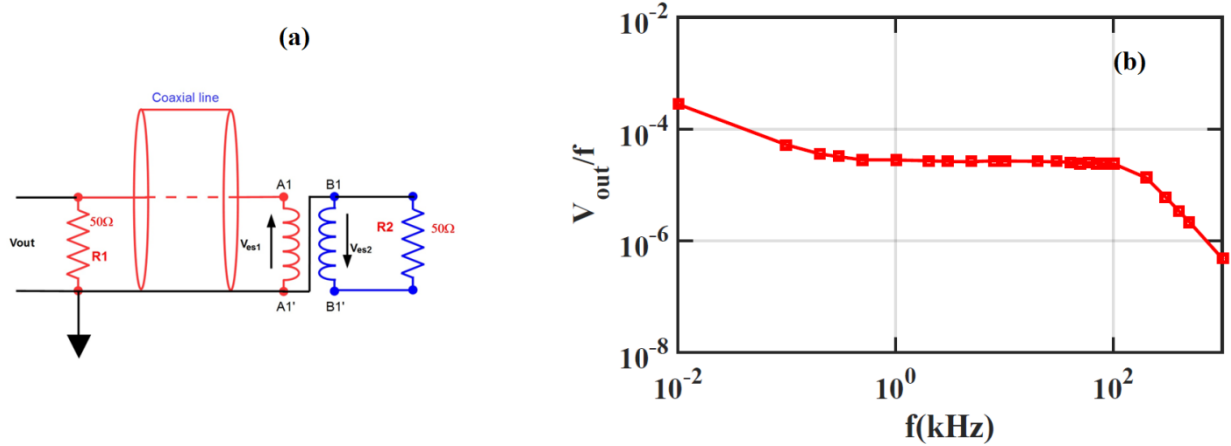

Figure 4: The electrical equivalent of the B-dot circuit is shown in (a). The probe eliminates capacitive effects by following the expression $\mathbf{V_{out}} = \mathbf{V_{A1A1'}} - \mathbf{V_{B1B1'}} = \mathbf{V_{ind1}} + \mathbf{V_{es1}} - (-\mathbf{V_{ind2}} + \mathbf{V_{es2}})$. Almost flat frequency response is observed in (b) for the probe between 1-100 kHz.

The probe is calibrated for inductive pickup. Figure (5) shows the pickup voltage when the probe is rotated between $0° - 180°$. Marginal asymmetry is observed between signals of $0^o$ and $180^0$, the reason for this may be due to the capacitive corruption. Here, zero is considered when the flux cutting surface is transverse to the field lines. The B-dot probe is calibrated for its response in plasma. The flux cutting surface of the probe rotated through $360^0$ to obtain a symmetric cosine waveform for pure inductive signal in plasma. A marginal deviation is again observed in the cosine waveform indicating minor poisoning due to capacitive pickup. The effect of the capacitive pickup is taken care in calculating the electromagnetic particle flux by the ratio of signals obtained at $0^0$ and $90^0$ with respect to $B_z$.

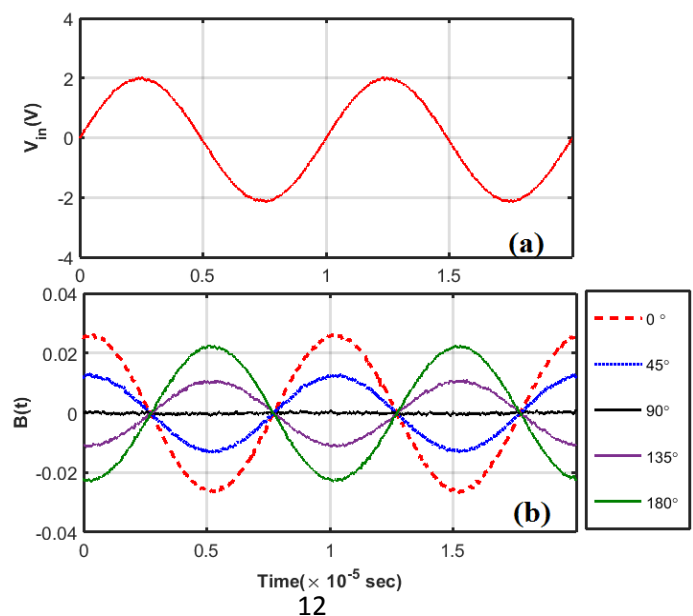

Figure 5: The input signal fed to the Helmholtz probe is shown in (a) and inductive responses of B-dot probe obtained for different orientations w.r.t. the Helmholtz magnetic field are shown in (b).

## III. Experimental Observations

### A. ETG Turbulence characterization

The plasma of the far target region from EEF is characterized for ETG threshold conditions[1]. We have divided the target plasma of LVPD in two regions namely, core ($x \leq 45\ cm$) and edge ($x > 45\ cm$) plasmas. The core plasma exhibits uniform plasma density profile but offers a finite gradient in electron temperature. It is important to mention here that EEF produces a transverse magnetic field of $B_x \sim 160G$ over its axial width of 4cm. The EEF magnetic field falls rapidly beyond its boundaries and attains a magnitude, equal to the ambient axial magnetic field of $B_z = 6.2G$ produced by a set of 10 garlanded coils.

The mean plasma parameters namely, plasma density, $n_e$, electron temperature, $T_e$, and floating potential, $V_f$, are measured using compensated cold Langmuir probes [40,41]. The plasma density is measured with probe biased to a fixed voltage of -80 V and its fluctuations are measured by picking its AC component. For electron temperature measurement, the probe is swept between -80V to 20V with sweep time, $\Delta t_{sweep} \sim 500\mu s$ for the probe current, $I_P$. The electron temperature is estimated from the Ie versus V$_B$ characteristic of the probe by using the expression, $T_e = [d\ln(I_e)/dV_B]^{-1}$ and $I_e (= I_p - I_s)$ , here $I_e$ and V$_B$ are the electron current and probe bias voltage respectively. The characteristic exhibits no signature of the presence of tail electrons in the region. However, temperature fluctuations are estimated by using a pair of closely separated Langmuir probes ($\phi_{P_{1,2}} \approx 1mm, L_{P_{1,2}} = 5mm$ and $\Delta x_{P_{1,2}} = 5mm$) from $\delta T_e = e(\phi_2 - \phi_1)/\ln(I_{e1}/I_{e2}))$ where $\phi_1, \phi_2$ and $I_{e1}, I_{e2}$ are the bias voltages and electron currents of the respective probes[1]. The probe orientation for all these measurements is always maintained perpendicular to the axial magnetic field.

The radial profiles are generated in the far target region of LVPD and are shown in figure 6. The plasma density and electron temperature gradient scale lengths are $L_n = [\frac{1}{n_e}(\frac{dn_e}{dx})]^{-1} \sim$ 300 cm and $L_T = [\frac{1}{T_e}(\frac{dT_e}{dx})]^{-1} \sim 50$ cm in the core region. In the outer region, the density fall becomes sharper while the electron temperature flattened out. Plasma potential does not show much variation in its value beyond an extent of $40 cm$. The threshold condition for ETG turbulence $\eta = L_n / L_T > 2/3$ is satisfied in the core region $x \leq 45 cm$ and an axial distance of $z \geq 70 cm$ away from the EEF. The maximum level of fluctuations reported for density, temperature and potential from the core region are 10 %, 30 %, and 1 % respectively[8].

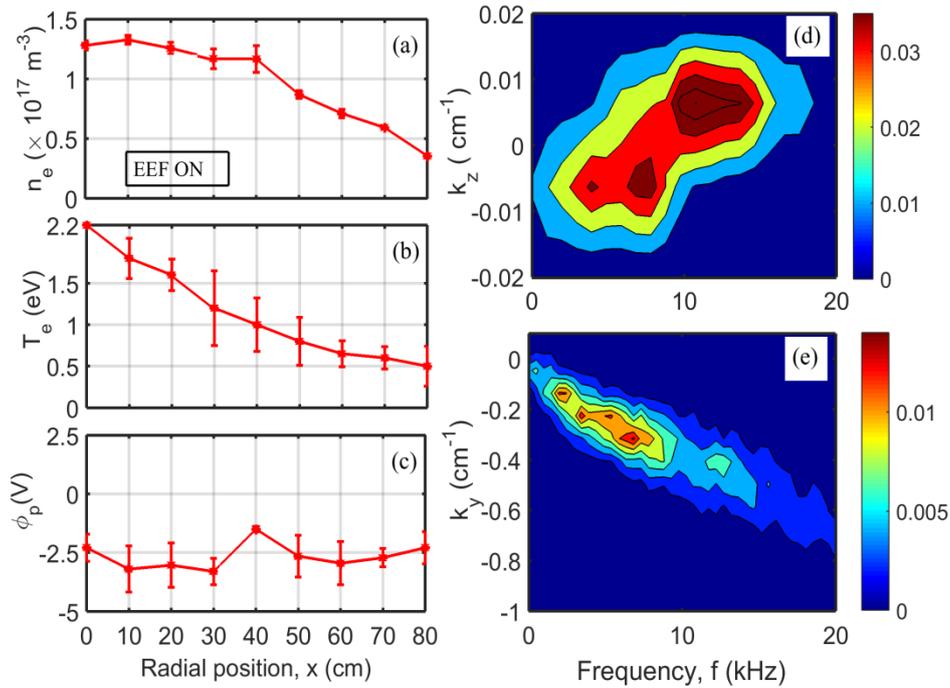

Figure 6: Radial equilibrium profiles for plasma parameters viz., (a) density, $n_e(m^{-3})$, (b) electron temperature, $T_e(eV)$, (c) plasma potential, $\phi_p$, are shown. The wave number frequency spectra, S(k, w) for parallel and perpendicular propagation are shown in figure (d) and (e) respectively.

The wave number- frequency spectra for parallel, $S(k_z, \omega)$ and poloidal, $S(k_y, \omega)$ propagation are measured to define characteristic features of the observed fluctuations in figure 6(d, e). The analysis for this is carried out following the work of Beall et al.,[42]. They are obtained by simultaneously measuring ion saturation current fluctuations with the pair of axially separated ($\Delta z \approx 30 cm$) and poloidally separated probes ($\Delta y = 1.5\ cm$). The large data length, $\geq 2 \times 10^5$ data points is used for the power spectral analysis. This is obtained from an ensemble of ~ 100 identical plasma discharges with sampling rate of 1MS/s. The data series is constructed by extracting 2048 data points from the steady state period of $6ms - 8ms$ from each plasma discharge. These data are segmented into 200 bins of 1024 data points each for obtaining a higher frequency resolution in the spectral analysis. The analysis exhibits confidence measure, $(1 - \frac{1}{\sqrt{N}}) * 100 = 93\ \%$, where $N$ is number of ensembles and variance, $\sigma^2 \sim \pm 0.01$ respectively. The measured maximum power from $S(k_z, \omega)$ for axial measurements correspond to frequency, $f \approx 7 kHz$ and wavelength, $|k_z| \approx 0.006 cm^{-1}$ respectively. Significant power for poloidal $S(k_y, \omega)$ resides in frequency band of $(25 - 90)\ krad/s$ in correspondence to wavenumber, $k_y \rho_e \sim 0.1 - 0.2$. These observations satisfy the frequency and wavelength ordering of ETG mode turbulence, $\Omega_{ci} < \omega(25 - 90)\ krad/s \ll \Omega_{ce}$, $k_\perp \rho_e \leq 1$ and $k_\perp \rho_i > 1$, where ($\Omega_{ci} \sim 2\ krad/s$) and $\Omega_{ce} \sim 94 \times 10^3 krad/s$) are ion and electron cyclotron frequency and $\rho_e (\sim 5mm)$ and $\rho_i (\sim 40 cm)$ are larmor radii of electron and ions respectively. The axial phase velocity $v_z \approx 7 \times 10^6 cm/s$ of the mode is two order higher than the perpendicular velocity, $v_\perp \approx 2 \times 10^4 cm/s$ [43].

A comparison of experimental observations is made with the estimates obtained from the generalized dispersion relation of ETG as suggested by Singh et al[30,43]. Singh et al., have shown that in the absence of magnetic gradient and curvature effects, similar to situations prevailing in linear devices, the long wave length "toroidal" ETG-like mode can be excited, because coupling of the slab ETG mode with the whistler mode at high β leads to the similar compression physics that is valid for toroidal mode. They have shown similarity in temperature perturbations produced in continuity equation by finite diamagnetic compressibility due to non-zero $\nabla_x B$ effect and finite diamagnetic compressibility due to $\delta B_z$ perturbation effect and for both cases, their responses to temperature perturbations, which are responsible for temperature gradient driven mode, emerge in same phase. Moreover, they have shown that W-ETG mode gets destabilized only when the electron temperature gradient crosses a threshold, $\eta_e > 2/3$.

The generalized dispersion relation for the ETG mode is expressed as

$$\hat{\omega}[\hat{\omega}\tau_e^* + \varepsilon_N \hat{k}_y + \hat{k}_\perp^2(\hat{\omega} - (\varepsilon_N + \varepsilon_T)\hat{k}_y) + \hat{\beta}(1+\tau_e^*)\{\hat{\omega} - (\varepsilon_N + \varepsilon_T)\hat{k}_y\}]$$
$$-\hat{\beta}[\hat{\omega} - (\varepsilon_N + \varepsilon_T)\hat{k}_y][(\varepsilon_T - 2\varepsilon_N/3)\hat{k}_y - 2\tau_e^*\hat{\omega}/3] \qquad (2)$$
$$= \hat{k}_\parallel^2 \hat{k}_\perp^2[(1+5\tau_e^*/3)\hat{\omega} - (\varepsilon_T - 2\varepsilon_N/3)\hat{k}_y]/[\hat{\omega}(\hat{\beta} + \hat{k}_\perp^2) - \hat{\beta}(\varepsilon_N + \varepsilon_T)\hat{k}_y]$$

Here we have introduced normalized parameters: $\hat{\beta} = \beta_e/2$, $\varepsilon_T = R/L_T$, $\varepsilon_n = R/L_n$, $\hat{\omega} = R\omega/c_e$, $\hat{k}_\perp = \hat{k}_y \sim \hat{k}_x = k_\perp \rho_e$, $\hat{k}_z = k_z R = k_\parallel R$, $\hat{\rho}_e = \rho_e/R$, $\tau_e = T_e/T_i$, and $\tau_e^* = \tau_e[1 - \tau_e \hat{\omega}^2 \hat{\rho}_e^2 m_i / \hat{k}_\perp^2 m_e]^{-1}$, where R is an arbitrary normalization length.

Figure (7) shows the growth rate, real frequency, and variation of normalized fluctuations in a magnetic field and electron temperature with plasma density for different plasma beta values. The figure 7(a, b) shows that a variation of linear growth rate ($\hat{\gamma}$) and real frequency ($\hat{\omega}$) with increasing values of beta, all other parameters used are shown in the figure itself.

Numerical predictions suggests that the growth rate for the observed mode peaks at $\hat{k}_y \sim 0.45$ in comparison to the work of Mattoo et al., where they have shown that it observed turbulence peaks at $\hat{k}_y \sim 0.2$. In pursuit of developing an understanding on ETG turbulence, Mattoo et al., have successfully dressed up plasma using transverse field diffusion across a large electron energy filter(EEF ) for no non- thermal electrons but having flat density, and sharp gradient in electron temperature) and have successfully excited ETG turbulence.

Following model equations[30], Mattoo et al., have estimated non linearly saturated fluctuation levels with numerical values. The amplitudes of density fluctuation $\tilde{n}$, parallel magnetic field fluctuation $\tilde{B}$ and electron temperature fluctuation $\tilde{T}$ are expressed in term of potential fluctuation $\tilde{\phi}$,

$$\left|\frac{\delta n_e}{n_e}\right| = \left|\tau_e^*\right| \left|\frac{e\delta\phi}{T_e}\right| \qquad (3)$$

$$\left|\frac{\delta B_z}{B_z}\right| = \frac{\beta_e}{2}\left[1 + \frac{5}{3}\tau_e^* - (\varepsilon_T - \tfrac{2}{3}\varepsilon_n)\frac{k_y \rho_e}{R\omega/c_e}\right] \left|\frac{e\delta\phi}{T_e}\right| \qquad (4)$$

and

$$\left|\frac{\delta T_e}{T_e}\right| = \left[(\varepsilon_T - \frac{2}{3}\varepsilon_n)\frac{k_y \rho_e}{R\omega/c_e} - \frac{2}{3}\tau_e^*\right] \left|\frac{e\delta\phi}{T_e}\right| \qquad (5)$$

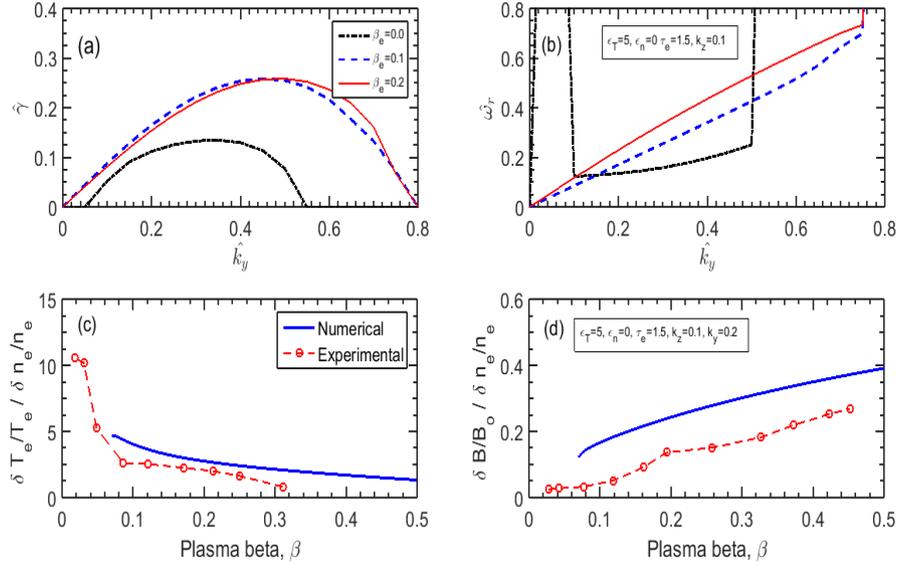

Figure 7: Normalized (a) growth rate $\hat{\gamma}$ and (b) real frequency, $\hat{\omega}_r$ verses $\hat{k}_\perp$ are shown for the ETG mode. A comparison of the ratio of numerically obtained normalised fluctuations of temperature and magnetic field with the normalised density fluctuations are shown in 7(c, d) along with experimental observations. The equations 3-5 are used for the estimation of normalized fluctuations in density, magnetic field and electron temperature respectively.

We have obtained growth rate and real frequency for the observed turbulence from the dispersion relation as shown in equation 2. Figure (7) shows the growth rate, real frequency, and variation of normalized fluctuations in a magnetic field and electron temperature with plasma density for different plasma beta values. The figure 7(a, b) shows that a variation of linear growth rate ($\hat{\gamma}$) and real frequency ($\hat{\omega}$) with increasing values of beta values. In figure 7(c, d), numerically obtained normalized value of fluctuations in temperature and magnetic field with density are compared with the experimentally obtained values. We examined the ratios of fluctuation amplitudes, as it eliminates the need for the absolute fluctuation amplitude of potential. We have obtained the ratios of fluctuations by using equations (3-5) for different plasma beta conditions imposed. For experimentally obtained $\hat{k}_y \sim 0.2$, the theoretically estimated values of, $\delta T_e/T / \delta n_e/n_e$ and $\delta B_z/B / \delta n_e/n_e$ agrees well with experimental observations.

## B. Characterization of $\delta J_{\parallel,e}$, $\delta J_{\parallel i}$ and Magnetic field Fluctuations

In finite beta plasma of LVPD, the observed ETG turbulence may exhibit characteristics of W-ETG turbulence, where parallel compression excited by the long wavelength of whistler wave gives additional perturbation in confining axial magnetic field [30]. The W-ETG turbulence due to finite $\beta$ effect causes magnetic fluctuations. By the use of bifilar B-dot probe, fluctuations in all the three components ($\delta B_x$, $\delta B_y$ and $\delta B_z$) of the magnetic field are measured.

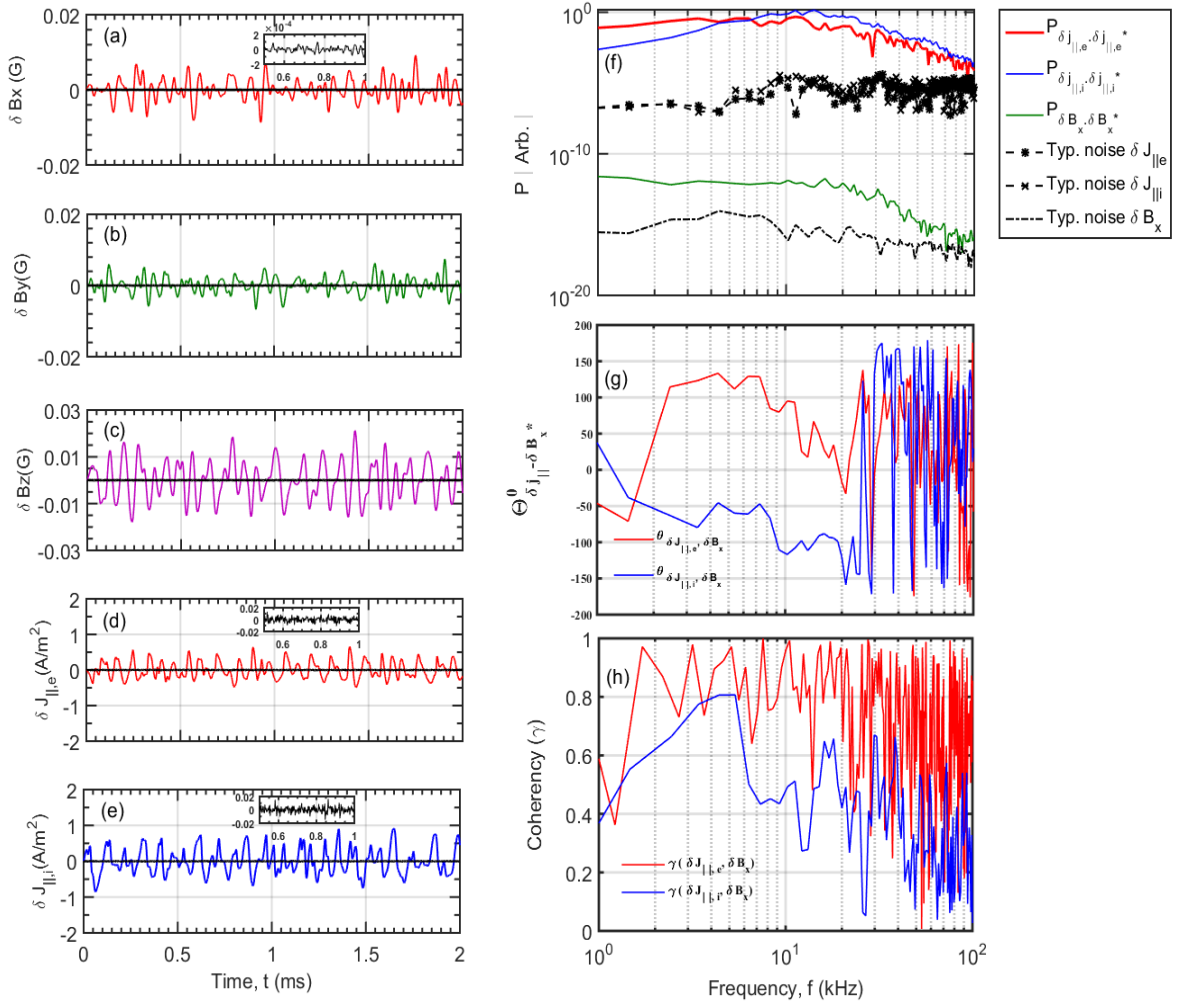

Figure 8: Time profile of magnetic field fluctuations in (a) x-component, $\delta B_x$ and the noise (b) y-component, $\delta B_y$ and (c) z-component, $\delta B_z$, and parallel current density fluctuations due to d) electrons, $J_{\parallel e}$ and e) ions, $\delta J_{\parallel i}$ and respective noise of the probe are shown. In figure (f-h), the auto power spectra of all signals and noise, phase angle and coherency between $\delta J_{\parallel e}$, $\delta J_{\parallel i}$ with $\delta B_x$ are shown.

Using specially designed directional probe arrangement, the parallel electron and ion current fluctuations are measured. Figure 8(a- e) shows a typical time series snap shot of fluctuations for all the three components of magnetic field and parallel electron and ion current respectively. The spectral characteristics are shown in Figure 8(f- h). The power spectra shows the spread of fluctuations over the frequency band of $1 - 20\ kHz$, in line with the $S(k_y, \omega)$ for W- ETG turbulence [figure 8(f)]. The phase angle of the fluctuations in radial magnetic field with electron(red) and ion(blue) currents, $\Theta_{\delta J_{\parallel e,i}, \delta B_x}$ remains uniform over the wide band of frequency [figure 8(g)]. The coherency, $\gamma(f)$ between the respective signals is high $\geq 0.6$ over the frequency band of 1- 10 kHz [figure 8(h)]. A comparison of the measured signal for both directional and B-dot probes with noise shows significant S/N ratio in measurements. The data acquisition system (PXI -NI5105, 12 bit digitization), resolves voltage signals $\sim 0.25 \mu V > noise(\sim 1 \mu V)$.

C. **Radial profiles of normalized fluctuation**

The radial profiles of normalized fluctuation in $B_x, B_y, B_z, J_{\parallel e}$ and $J_{\parallel i}$ are measured and are shown in figure (9). We observed that normalized fluctuation amplitudes are higher in the core region and the typical levels of $\delta B_x$, $\delta B_y$ and $\delta B_z$ lies between $0.05\ \% - 0.2\ \%$ whereas, the normalized parallel current fluctuations because of electron and ions are 10 % to 20 %.

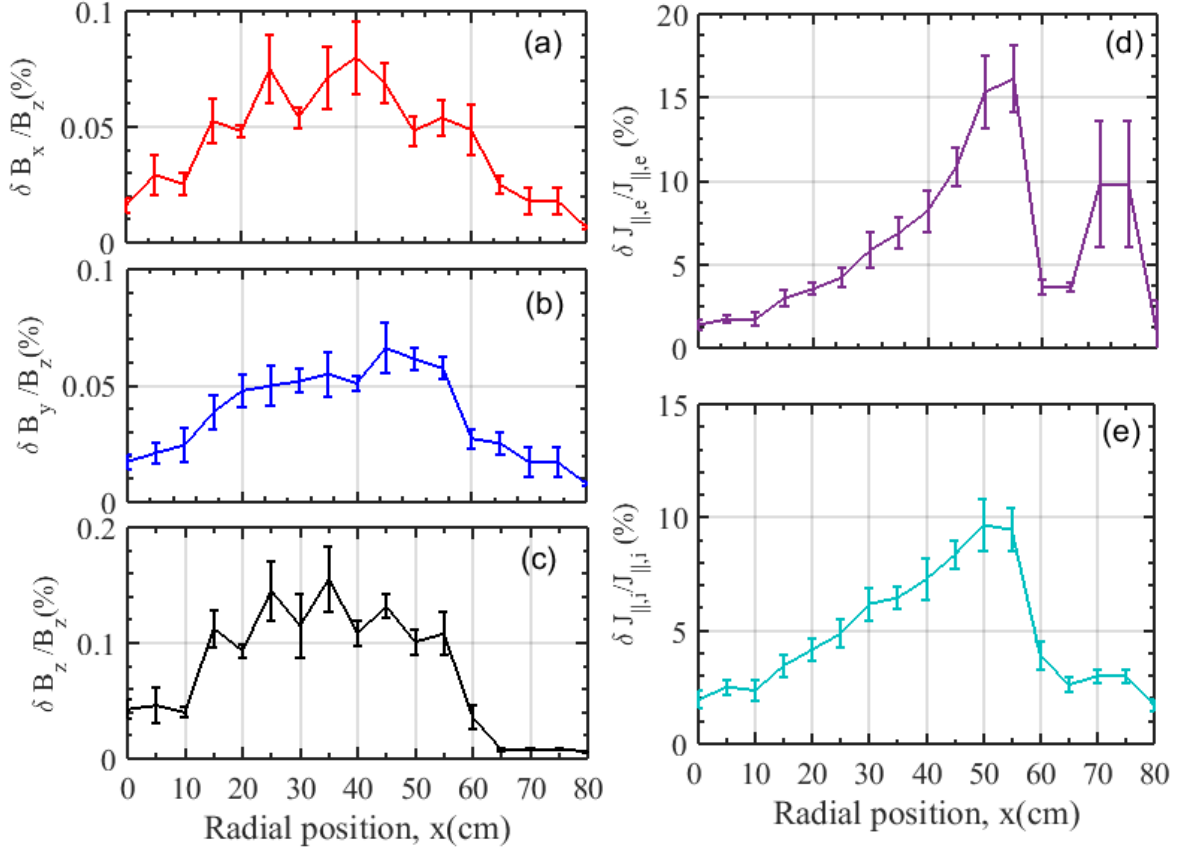

Figure 9: Typical radial profiles of normalized fluctuations in three components of the magnetic field and parallel electron and ion current are shown in (a-c) and (d-e) respectively. The level maximizes radially, between 20- 50 cm.

**D. EM flux measurement and comparison with electrostatic particle flux**

Both electrostatic and electromagnetic particle fluxes due to fluctuations are measured along the radius of the target chamber. The electrostatic particle flux ($\Gamma_{es} = \delta n_e \delta v_r$) is measured from the correlated density ($\delta n_e$) and radial velocity ($\delta v_x$) fluctuations. The velocity fluctuations are estimated from the cross product of fluctuating poloidal electric field ($\delta E_y$) and ambient magnetic field, $\delta E \times B$, where $\delta E$ is estimated from the floating potential fluctuations measured by a pair of poloidally separated Langmuir probes[8]. The electromagnetic flux derived from $\Gamma_{em} = \frac{1}{qB} \delta J_{\parallel q} \delta B_x$, where 'q' is the charge, is observed to be $\sim 10^{-5} \times <\Gamma_{es}>$ but is finite.

Figure (10) shows a comparison of the estimated electromagnetic particle flux due to electron current with electrostatic flux at x=30 cm in the target chamber.

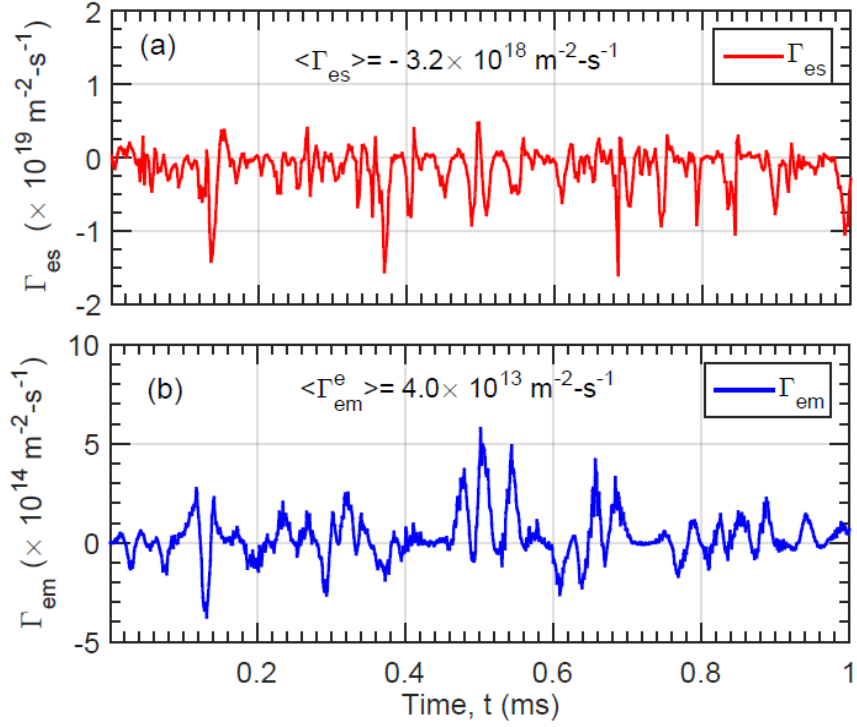

Figure 10: The time series of simultaneously measured electrostatic (a) and electromagnetic (b) fluxes in ETG background at x=30 cm are shown. The typical time-averaged values obtained are $<\Gamma_{es}> = -3.2 \times 10^{18} m^{-2} s^{-1}$ and $<\Gamma_{em}> = 4 \times 10^{13} m^{-2} s^{-1}$ respectively.

We found that the electromagnetic flux is finite even though as mentioned, it should be zero[44]. This has prompted us to look for the reason that why EM flux is finite in ETG plasma in slab geometry of LVPD.

## IV. THEORETICAL MODEL AND EXPERIMENTAL COMPARISONS

### A. Quasilinear theory for electromagnetic particle flux

Electromagnetic electron particle flux is given by,

$$\Gamma_{em}^e = \frac{-1}{eB} < \delta J_{\parallel,e} \delta B_r > \qquad (6)$$

Since the total current perturbation $\delta J_\parallel$ is sum of electron and ion current perturbations $\delta J_{\parallel i} + \delta J_{\parallel e} = \delta J_\parallel$, the electron flux can be written as

$$\Gamma_{em}^e = \frac{1}{eB} < \delta J_{\parallel i} \delta B_x > - \frac{1}{eB} < \delta J_\parallel \delta B_x > \qquad (7)$$

Then noting that $\delta J_\parallel = \frac{c}{4\pi} (\nabla \times \delta B)_{z-component}$. Here, the displacement current contribution is ignored as $\frac{1}{c^2}\frac{\partial E}{\partial t}$ scales with $(L/c^2T^2)B$, which is negligible. Here, L is characteristic length, T is characteristic time scale and c is the speed of light.

$$\delta J_\| \delta B_x = \frac{c}{4\pi}[\frac{\partial}{\partial x}(\delta B_x \delta B_y) - \delta B_y \frac{\partial}{\partial x}\delta B_x - \frac{\partial}{\partial y}(\frac{1}{2}\delta B_x^2)]$$

Further using $\nabla.B = 0 \Rightarrow \frac{\partial}{\partial x}\delta B_x = -(\frac{\partial}{\partial y}\delta B_y + \frac{\partial}{\partial z}\delta B_z)$ and hence

$$<\delta J_\| \delta B_x> = \frac{c}{4\pi}[\frac{\partial}{\partial x}(<\delta B_x \delta B_y>) + \frac{\partial}{\partial y}<\frac{1}{2}\delta B_y^2> + <\delta B_y \frac{\partial}{\partial z}\delta B_z> - \frac{\partial}{\partial y}<\frac{1}{2}\delta B_x^2>]$$

The terms $\frac{\partial}{\partial y}<\frac{1}{2}\delta B_y^2>$ and $\frac{\partial}{\partial y}<\frac{1}{2}\delta B_x^2>$ vanishes due to poloidal averaging. Poloidal derivative of poloidally averaged quantity is zero. So eventually the expression for flux becomes

$$\Gamma_{em}^e = \frac{1}{eB}<\delta J_{\|i}\delta B_x> - \frac{1}{eB}\frac{c}{4\pi}\frac{\partial}{\partial x}<\delta B_x\delta B_y> - \frac{1}{eB}\frac{c}{4\pi}<\delta B_y\frac{\partial}{\partial z}\delta B_z> \qquad (8)$$

The first term on the right hand side though looks like an electromagnetic ion flux is not a true ion flux. Because of ETG mode frequency scaling, $\Omega_{ci} < \omega \ll \Omega_{ce}$ the ions are unmagnetized and hence the electromagnetic radial ion velocity perturbation is not same as the electron radial velocity perturbation. In fact it can be shown that the radial ion drift velocity perturbation is zero at the leading order. Hence, the actual turbulent ion flux is zero when ions are unmagnetized in ETG turbulence. The first term had been actual ion flux in the frequency range $\omega \ll \Omega_{ci}$ (e.g. ITG turbulence) when ions are magnetized and the ions radial drift velocity fluctuations is given by the same expression as for the electrons to the leading order in $\Omega_{ci}$. Hence the first term represents a pseudo ion flux in ETG turbulence due to its morphological similarity with the actual ion flux in ITG turbulence. The second term is the divergence of Maxwell-stress $<\delta B_x\delta B_y>$ and third term is another correlation between $\delta B_y$ and $\frac{\partial}{\partial z}\delta B_z$. Equation (8) can also be expressed in terms of the vector potentials as follows

$$\Gamma_{em}^e = \frac{1}{eB}<\delta J_{\|i}\delta B_x> - \frac{c}{4\pi}\frac{1}{B}\nabla_x<\frac{\partial}{\partial x}\delta A_z\frac{\partial}{\partial z}\delta A_y> + \frac{c}{4\pi}\frac{1}{B}<\vec{\nabla}\delta A_z.\vec{\nabla}\frac{\partial}{\partial z}\delta A_y> \qquad (9)$$

This clearly shows that the limit of $\delta A_y \to 0$ the electron flux is purely made of pseudo ion flux due to parallel ion current fluctuations, while the conventional theory by Holland and Diamond et al. [44] produces no flux. This is the new significant improvement in the understanding of electromagnetic flux in ETG turbulence. We next obtained a quasilinear expression for the ion flux. From the parallel ion momentum equation it is straight forward, to arrive at the following equation for parallel current perturbation. Assuming quasi neutrality, no collisionality and cold ions

$$\delta J_{\|,i} = \frac{e^2}{m_i} n \left(\frac{1}{\omega} k_\| \delta\phi - \frac{1}{c} A_\|\right) - \frac{e}{m_i}\left(-\frac{1}{\omega} k_\| \delta p_i + \frac{i}{\omega}\frac{\delta B_r}{B} P'_{i0}\right) \quad (10)$$

Here $\delta p_i$ is ion pressure perturbation and $\grave{P}_{io}$ is equilibrium ion pressure gradient. Hence electromagnetic pseudo ion flux becomes

$$\Gamma^i_{em} = \frac{1}{eB} <\delta J_{\|i}\delta B_r>$$

$$= \sum_{\vec{k}}\left[\frac{e}{m_i B} n\left(\frac{1}{\omega} k_\| \delta\phi - \frac{1}{c} A_\|\right) - \frac{1}{m_i B}\left(-\frac{1}{\omega} k_\| \delta p_i + \frac{i}{\omega}\frac{\delta B_r}{B} P'_{i0}\right)\right](-ik_y A_\|^*)$$

$$= \sum_{\vec{k}}\left[\frac{e}{m_i B} n\left(-i\frac{1}{\omega} k_\| k_y \delta\phi A_\|^* + i\frac{1}{c}k_y|A_\||^2\right) - \frac{1}{m_i B}\left(i\frac{1}{\omega} k_\| k_y \delta p_i A_\|^* + i\frac{1}{\omega}k_y^2 \frac{1}{B}|A_\||^2 P'_{i0}\right)\right]$$

$$= \sum_{\vec{k}}\frac{\beta_e}{2}\frac{e}{m_i B} n \frac{k_\| k_y}{|\omega|^2}\left(-\omega_r Im(R_A) + \gamma Real(R_A)\right)|\delta\phi|^2 - \sum_{\vec{k}}\frac{1}{m_i B}\left(i\frac{1}{\omega} k_\| k_y \delta p_i A_\|^* + \frac{\gamma}{|\omega|^2}k_y^2 \frac{1}{B}|A_\||^2 P'_{io}\right)$$

$$= \Gamma^i_{em1} + \Gamma^i_{em2} + \Gamma^i_{em3}(say) \quad (11)$$

where the linear electromagnetic response function is given by

$$R_A = \frac{k_z\omega - \frac{5\tau_e}{3}\omega - \left(\eta_e - \frac{2}{3}\right)k_y}{\omega\left(\frac{\beta_e}{2}+k_\perp^2\right)\omega - \frac{\beta_e}{2}Kk_y}$$

It is more enlightening to analyze the different pieces of the flux separately

$$\Gamma_{em1} = \sum_{\vec{k}}\frac{\beta_e}{2}\frac{m_e}{m_i} nc_e \frac{k_\| c_e k_y \rho_e}{|\omega|^2}\left(-\omega_r Im(R_A) + \gamma Real(R_A)\right)\left|\frac{e\delta\phi}{T_e}\right|^2 \quad (12)$$

$$\Gamma_{em2} = -\sum_{\vec{k}}\tau_i \frac{\beta_e}{2}\frac{m_e}{m_i} nc_e i\frac{1}{\omega}k_\| c_e k_y \rho_e R_A^*\left(\frac{\delta n}{n_0} + \frac{\delta T_i}{T_i}\right)\left(\frac{e\delta\phi^*}{T_{e0}}\right)$$

$$= \sum_{\vec{k}}\frac{\beta_e}{2}\frac{m_e}{m_i} nc_e \frac{k_\| c_e k_y \rho_e}{|\omega|^2}\left[(\omega_r - \gamma\delta_k)Im(R_A) - (\gamma + \omega_r\delta_k)Real(R_A)\right]\left|\frac{e\delta\phi}{T_e}\right|^2 + \alpha\tau_i \frac{\delta T_i}{T_{i0}}\frac{e\delta\phi^*}{T_{eo}} \quad (13)$$

Where we used the non-Boltzmannian ion response coming from the resonance of the ETG mode with the ions[8,45]

$$\frac{\delta n_i}{n_0} = -\tau_e(1 + i\delta_k)\frac{e\delta\phi_k}{T_{e0}} \quad (14)$$

Where the non-adiabatic parameter $\delta_k$ is given by $\delta_k = \sqrt{\pi}\frac{\omega}{k_\perp c_i}\exp-\left(\frac{\omega}{k_\perp c_i}\right)^2$.

The third piece is

$$\Gamma_{em3} = \sum_{\vec{k}} \tau_i \left(\frac{\beta_e}{2}\right)^2 \frac{m_e}{m_i} nc_e \frac{\gamma c_e L_{pi}^{-1}}{|\omega|^2} k_y^2 \rho_e^2 |R_A|^2 \left|\frac{e\delta\phi}{T_{e0}}\right|^2 \qquad (15)$$

Here $L_{pi}^{-1}$ is pressure gradient scale length. It is noticeable that $\Gamma_{em3}$ is of higher order in $\frac{\beta_e}{2}$ as compared to $\Gamma_{em1}$ and $\Gamma_{em2}$. Assuming $\frac{\delta T_i}{T_{i0}} = 0$ and noting that part of $\Gamma_{em2}$ cancels with $\Gamma_{em1}$, we get the total radial flux as

$$\Gamma_{em}^i = -\frac{\beta_e}{2}\frac{m_e}{m_i} n_0 c_e \sum_{\vec{k}} \frac{k_\parallel c_e k_y \rho_e}{|\omega|^2} \delta_k [(\gamma Im(R_A) + \omega_r Real(R_A))] \left|\frac{e\delta\phi_k}{T_{e0}}\right|^2 + O\left(\left(\frac{\beta_e}{2}\right)^2 \frac{m_e}{m_i}\right) \qquad (16)$$

Now recall the expression for electrostatic electron flux $\Gamma_{es}$ from Ref[8].

$$\Gamma_{es} = <\delta n_e \delta v_{E\times B}> = \sum_k n_o c_e \tau_e \delta_k k_y \rho_e \left|\frac{e\delta\phi_k}{T_{eo}}\right|^2 \qquad (17)$$

For the typical $k_z = 0.1$, $\omega = 0.1(1+i)$, $k_y = 0.1$, $\tau_e = 10$, $\eta_e = 3$, $\beta = 0.2$, the ratio of electromagnetic to electrostatic flux yields

$$\frac{\Gamma_{em}^i}{\Gamma_{es}} \approx \frac{\beta_e}{2}\frac{m_e}{m_i}\frac{1}{\tau_e} \times (10 \sim 100) = 0.08 \times \frac{1}{40\times 1836} \times \frac{1}{\tau_e}(10 \sim 100) \approx 10^{-5} \qquad (18)$$

,

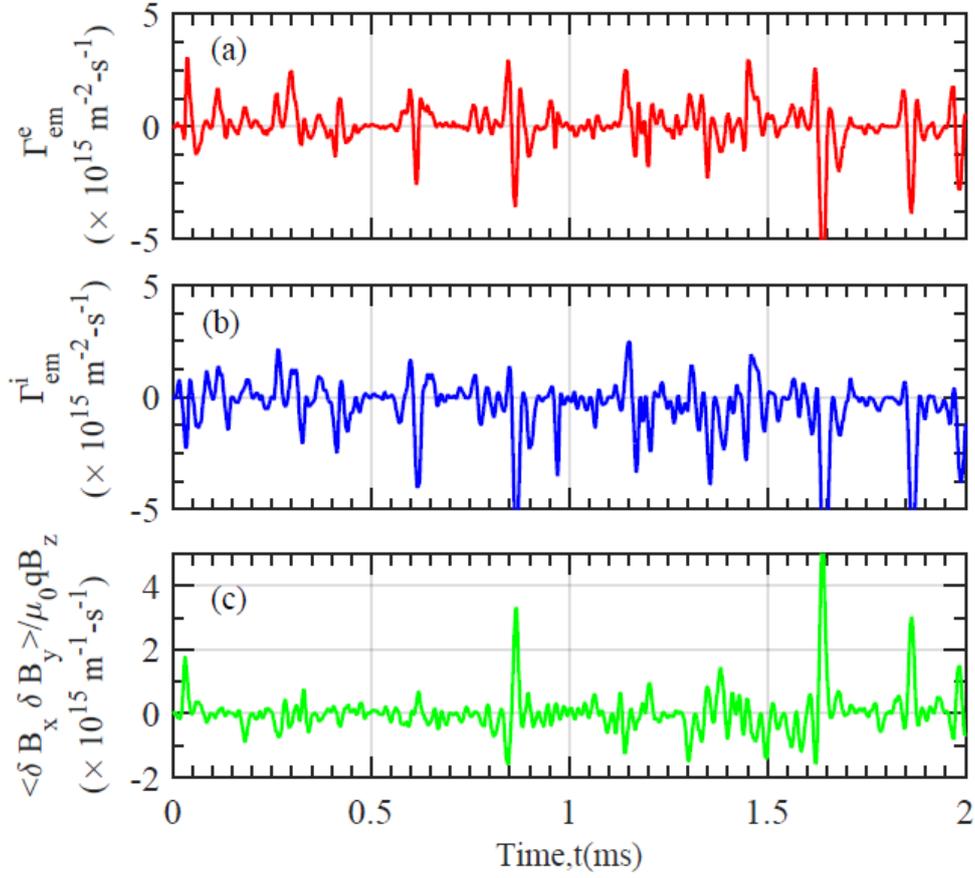

Figure 11: Electromagnetic particle flux is shown due to parallel (a) electron streaming, $\Gamma^e_{em}$, (b) ion streaming, $\Gamma^i_{em}$ and (c) time variation of Maxwell stress at x=30 cm. The average values of the flux components due to parallel electron and ion currents are $\mathbf{4.0 \times 10^{13} m^{-2} s^{-1}}$ and $\mathbf{-1.2 \times 10^{14} m^{-2} s^{-1}}$ respectively and the Maxwell Stress contribution is about $\mathbf{-3.2 \times 10^{13} m^{-1} s^{-1}}$.

In figure 11, we have made an attempt to estimate the experimentally observed particle flux due to electron and ion parallel motion, It was observed that the average electron and ion flux contribution differ significantly. This suggested that the electromagnetic flux may have a non ambipolar nature which can probably leads to the creation of radial charge separation balanced by the sum of divergence of Maxwell –stress and grad of axial magnetic field fluctuations.

As can be seen from equation (5), the difference in flux produced by electron and ions can be understood by the radial evolution of Maxwell stress.

### B. Radial profiles of ion and electron fluxes and comparison with theoretically obtained values

In figure 12, the electromagnetic fluctuations driven radial electron flux is measured during the steady state period between (6- 8) ms of discharge pulse using the correlated parallel electron and ions current density fluctuations, $\delta J_{\parallel e,i}$ and radial magnetic field fluctuations, $\delta B_x$. The contribution to flux due to ion current is compared with the theoretically obtained values. The theoretical values are obtained using the expression

$$\Gamma_{em}^i = -\frac{\beta_e}{2}\frac{m_e}{m_i} n_0 c_e \sum_{\vec{k}} \frac{k_\parallel c_e k_y \rho_e}{|\omega|^2} \delta_k [(\gamma Im(R_A) + \omega_r Real(R_A))] \left|\frac{e\delta\phi}{T_{e0}}\right|^2 + O\left(\left(\frac{\beta_e}{2}\right)^2 \frac{m_e}{m_i}\right)$$

By the use of experimentally measured parameters such as parallel wave number, $|k_\parallel| \sim 0.006\ cm^{-1}$, poloidal wave number, $|k_y| \approx 0.15 cm^{-1}$ (from figure 6), local experimental $\beta_e$ and local fluctuation levels, $\left|\frac{e\delta\phi}{T_e}\right|$, we found that the estimated value of flux agrees well with the experimentally obtained ion current flux.

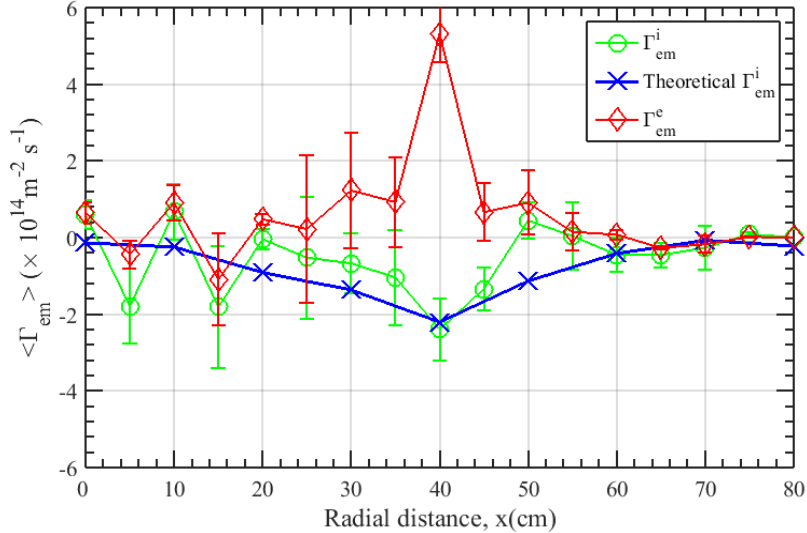

Figure 12: Radial electromagnetic particle flux due to parallel electron current, $\Gamma_{\delta J_{\parallel e}\delta B_x}$, parallel ion streaming current, $\Gamma_{\delta J_{\parallel i}\delta B_x}$ and numerical values obtained from theoretical the expression for $\Gamma_{em}^i$.

## V. Summary and Conclusion

We studied electromagnetic particle flux due to ETG turbulence in LVPD. The measurement shows that the electromagnetic flux is non-ambipolar leading to the net charge flux. The electromagnetic flux is found to be smaller than the electrostatic flux. A quasilinear theory is proposed to explain the observed flux which convinces that the non-ambipolarity is due to divergence of Maxwell stress $\frac{\partial}{\partial x}<\delta B_x \delta B_y>$ and another correlation $<\delta B_y \frac{\partial}{\partial z}\delta B_z>$. The pseudo ion flux similar to ion flux in morphology had the ions been magnetized. This is because the parallel ion dynamics is little affected by the degree of magnetization of ions. The physical electromagnetic ion flux in ETG turbulence is zero because no radial ion drift fluctuation due to unmagnetization of ions. Then the overall quasi-neutrality must be maintained by the parallel fluxes. The quasilinear estimates of the electromagnetic pseudo ion flux due to parallel ion current fluctuation is compared well with measured flux. The complete study of electromagnetic electron flux also requires measurement of divergence of Maxwell-stress which together with the Reynolds stress, $<\delta v_x \delta v_y>$ may be responsible for the $E \times B$ shear layer formation at $x = 40\ cm$ (evident from the pick of the plasma potential in Figure 6(c)).This is still an on-going work and will be presented elsewhere.

## Acknowledgment

The authors are greatly thankful to Mr. Dharmik Trivedi of LVPD group for his technical help. Authors express their special thanks to Dr. Nirmal Bisai for providing extremely useful comments while reviewing the paper. One of the authors R. Singh acknowledges for support by the R&D Program through National Fusion Research Institute (NFRI) funded by the Ministry of Science, ICT and Future Planning of the Republic of Korea (NFRI-EN1541-2), and by the same Ministry under the ITER technology R&D.

## References

[1]   Mattoo S K, Singh S K, Awasthi L M, Singh R and Kaw P K 2012 Experimental observation of electron-temperature-gradient turbulence in a laboratory plasma *Phys. Rev. Lett.* **108** 1–5

[2]   Liewer P C 1985 Measurements of microturbulence in tokamaks and comparisons with theories of turbulence and anomalous transport *Nucl. Fusion* **25** 543–621

[3]   Wootton A J, Carreras B A, Matsumoto H, McGuire K, Peebles W A, Ritz C P, Terry P W and Zweben S J 1990 Fluctuations and anomalous transport in tokamaks *Phys. Fluids B* **2** 2879–903

[4]   Nielsen A H, Pécseli H L and Rasmussen J J 1995 Electrostatic fluctuations and turbulent-plasma transport in low-beta plasmas *Phys. Scr.* **51** 632–7

[5]   Fasoli A, Burckel A, Federspiel L, Furno I, Gustafson K, Iraji D, Labit B, Loizu J, Plyushchev G, Ricci P, Theiler C, Diallo A, Mueller S H, Podestà M and Poli F 2010 Electrostatic instabilities, turbulence and fast ion interactions in the TORPEX device *Plasma Phys. Control. Fusion* **52** 124020

[6]     Fasoli A, Labit B, McGrath M, Müller S H, Plyushchev G, Podestà M and Poli F M 2006 Electrostatic turbulence and transport in a simple magnetized plasma *Phys. Plasmas* **13**

[7]     Garbet X, Garzotti L, Mantica P, Nordman H, Valovic M, Weisen H and Angioni C 2003 Turbulent Particle Transport in Magnetized Plasmas *Phys. Rev. Lett.* **91** 35001

[8]     Srivastav P, Singh R, Awasthi L M, Sanyasi A K, Srivastava P K, Sugandhi R, Singh R and Kaw P K 2017 Observation of radially inward turbulent particle flux in ETG dominated plasma of LVPD *Phys. Plasmas* **24** 112115

[9]     Kim J Y and Horton W 1991 Transition from toroidal to slab temperature gradient driven modes *Phys. Fluids B* **3** 1167–70

[10]    Callen J D, Groebner R J, Osborne T H, Canik J M, Owen. L W, Pankin A Y, Rafiq T, Rognlien T D and Stacey W M 2010 Analysis of pedestal plasma transport *Nucl. Fusion* **50**

[11]    Castejon F and Team the T-I 2016 3D effect on transport and plasma control in the TJ-II stellarator *Nucl. Fusion*

[12]    Ding W X, Brower D L, Craig D, Deng B H, Prager S C, Sarff J S and Svidzinski V 2007 Nonambipolar magnetic-fluctuation-induced particle transport and plasma flow in the MST reversed-field pinch *Phys. Rev. Lett.* **99** 1–4

[13]    Rempel T D, Almagri A F, Assadi S, Den Hartog D J, Hokin S A, Prager S C, Sarff J S, Shen W, Sidikman K L, Spragins C W, Sprott J C, Stoneking M R and Zita E J 1992 Turbulent transport in the Madison Symmetric Torus reversed-field pinch *Phys. Fluids B* **4** 2136–41

[14]    Stoneking M R, Hokin S A, Prager S C, Fiksel G, Ji H and Den Hartog D J 1994 Particle Transport Due to Magnetic Fluctuations *Phys. Rev. Lett.* **73** 549–52

[15]    Ryter F, Leuterer F, Pereverzev G, Fahrbach H U, Stober J and Suttrop W 2001 Experimental evidence for gradient length-driven electron transport in tokamaks *Phys. Rev. Lett.* **86** 2325–8

[16]    Hoang G T, Bourdelle C, Garbet X, Giruzzi G, Aniel T, Ottaviani M, Horton W, Zhu P and Budny R V. 2001 Experimental Determination of Critical Threshold in Electron Transport on Tore Supra *Phys. Rev. Lett.* **87** 125001

[17]    Horton W, Wong H V., Morrison P J, Wurm A, Kim J H, Perez J C, Pratt J, Hoang G T, LeBlanc B P and Ball R 2005 Temperature gradient driven electron transport in NSTX and Tore Supra *Nucl. Fusion* **45** 976–85

[18]    Kaye S M, Levinton F M, Stutman D, Tritz K, Yuh H, Bell M G, Bell R E, Domier C W, Gates D, Horton W, Kim J, LeBlanc B P, Luhmann N C, Maingi R, Mazzucato E, Menard J E, Mikkelsen D, Mueller D, Park H, Rewoldt G, Sabbagh S A, Smith D R and Wang W 2007 Confinement and local transport in the National Spherical Torus Experiment (NSTX) *Nucl. Fusion* **47** 499–509

[19]    Fu X R, Horton W, Xiao Y, Lin Z, Sen A K and Sokolov V 2012 Validation of electron temperature gradient turbulence in the Columbia Linear Machine *Phys. Plasmas* **19**

[20]    Moon C, Kaneko T and Hatakeyama R 2013 Dynamics of nonlinear coupling between electron-temperature-gradient mode and drift-wave mode in linear magnetized plasmas

*Phys. Rev. Lett.* **111** 1–4

[21]   Callen J D 1977 Drift-wave turbulence effects on magnetic structure and plasma transport in tokamaks *Phys Rev Lett* **39** 1540–3

[22]   Rechester A B and Rosenbluth M N 1978 Electron heat transport in a tokamak with destroyed magnetic surfaces *Phys. Rev. Lett.* **40** 38–41

[23]   Hutchinson I H, Malacarne M, Noonan P and Brotherton-Ratcliffe D 1984 The structure of magnetic fluctuations in the HBTX-1A reversed field pinch *Nucl. Fusion* **24**

[24]   Carreras B A, Diamond P H, Murakami M, Dunlap J L, Bell J D, Hicks H R, Holmes J A, Lazarus E A, Paré V K, Similon P, Thomas C E and Wieland R M 1983 Transport effects induced by resistive ballooning modes and comparison with high-$\beta p$ ISX-B tokamak confinement *Phys. Rev. Lett.* **50** 503–6

[25]   Ohyabu N, Jahns G L, Stambaugh R D and Strait E J 1987 Correlation of Magnetic Fluctuations and Edge Transport in the Doublet HI Tokamak **2** 120–3

[26]   García-Muñoz M, Hicks N, van Voornveld R, Classen I G J, Bilato R, Bobkov V, Bruedgam M, Fahrbach H-U, Igochine V, Jaemsae S, Maraschek M and Sassenberg K 2010 Convective and Diffusive Energetic Particle Losses Induced by Shear Alfvén Waves in the ASDEX Upgrade Tokamak *Phys. Rev. Lett.* **104** 185002

[27]   Stoneking M R 1994 Particle transport due to magnetic fluctuations *Phys. Rev. ...* **73** 549–53

[28]   Shen W, Dexter R N and Prager S C 1992 Current-density fluctuations and ambipolarity of transport *Phys. Rev. Lett.* **68** 1319–22

[29]   Ding W X, Brower D L, Fiksel G, Den Hartog D J, Prager S C and Sarff J S 2009 Magnetic-Fluctuation-Induced Particle Transport and Density Relaxation in a High-Temperature Plasma *Phys. Rev. Lett.* **103** 1–4

[30]   Singh S K, Awasthi L M, Singh R, Kaw P K, Jha R and Mattoo S K 2011 Theory of coupled whistler-electron temperature gradient mode in high beta plasma: Application to linear plasma device *Phys. Plasmas* **18**

[31]   Holland C and Diamond P H 2002 Electromagnetic secondary instabilities in electron temperature gradient turbulence *Phys. Plasmas* **9** 3857–66

[32]   Mattoo S K, Anitha V P, Awasthi L M, Ravi G, Srivastava P, Kumar K, Kumar T A S, Rajyaguru C, Doshi B, Baruah U, Patel G B, Patel P and Buch B N 2001 A large volume plasma device *Rev. Sci. Instrum.* **72** 3864–72

[33]   Hudis M and Lidsky L M 1970 Directional Langmuir probe *J. Appl. Phys.* **41** 5011–7

[34]   Dejarnac R, Gunn J P, Stöckel J, Adámek J, Brotánková J and Ionita C 2007 Study of ion sheath expansion and anisotropy of the electron parallel energy distribution in the CASTOR tokamak *Plasma Phys. Control. Fusion* **49** 1791–808

[35]   Awasthi L M, Mattoo S K, Jha R, Singh R and Kaw P K 2010 Study of electromagnetic fluctuations in high beta plasma of a large linear device *Phys. Plasmas* **17** 1–12

[36]   Singh S K, Srivastava P K, Awasthi L M, Mattoo S K, Sanyasi A K, Singh R and Kaw P K 2014 Performance of large electron energy filter in large volume plasma device *Rev. Sci. Instrum.* **85**

[37]  Sanyasi A K, Awasthi L M, Mattoo S K, Srivastava P K, Singh S K, Singh R and Kaw P K 2013 Plasma response to electron energy filter in large volume plasma device *Phys. Plasmas* **20** 122113

[38]  Sanyasi A K, Srivastava P K and Awasthi L M 2017 Plasma potential measurement using centre tapped emissive probe (CTEP) in laboratory plasma *Meas. Sci. Technol.* **28** 45904

[39]  Nagaoka K, Okamoto A, Yoshimura S and Tanaka M Y 2001 Plasma flow measurement using directional Langmuir probe under weakly ion-magnetized conditions *J. Phys. Soc. Japan* **70** 131–7

[40]  Chung K S 2012 Mach probes *Plasma Sources Sci. Technol.* **21** 63001

[41]  Everson E T, Pribyl P, Constantin C G, Zylstra A, Schaeffer D, Kugland N L and Niemann C 2009 Design, construction, and calibration of a three-axis, high-frequency magnetic probe (B-dot probe) as a diagnostic for exploding plasmas *Rev. Sci. Instrum.* **80**

[42]  Beall J M, Kim Y C and Powers E J 1982 Estimation of wavenumber and frequency spectra using fixed probe pairs *J. Appl. Phys.* **53** 3933–40

[43]  Singh S K, Awasthi L M, Mattoo S K, Srivastava P K, Singh R and Kaw P K 2012 Investigations on ETG turbulence in finite beta plasmas of LVPD *Plasma Phys. Control. Fusion* **54** 124015

[44]  Holland C and Diamond P H 2002 Electromagnetic secondary instabilities in electron temperature gradient turbulence *Phys. Plasmas* **9** 3857

[45]  Singh R, Nordman H, Anderson J and Weiland J 1998 Thermal transport in collision dominated edge plasmas *Phys. Plasmas* **5** 3669–74